\documentclass{article}
\usepackage{fa2020}
\usepackage{amsmath}
\usepackage{amssymb}
\usepackage{subcaption}
\usepackage{multicol} 
\usepackage{caption}
\usepackage{cite}
\usepackage{url}
\usepackage{graphicx}
\usepackage{color}
\usepackage[binary-units=true]{siunitx}

\title{wav2shape: Hearing the shape of a drum machine}

\twoauthors
  {Han Han} {Integrated Digital Media (IDM) Program \\ New York University}
  {Vincent Lostanlen} {Music and Audio Research Lab (MARL) \\ New York University}

\begin{document}

\maketitle
\begin{abstract}
Disentangling and recovering physical attributes, such as shape and material, from a few waveform examples is a challenging inverse problem in audio signal processing, with numerous applications in musical acoustics as well as structural engineering.
We propose to address this problem via a combination of time--frequency analysis and supervised machine learning.
We start by synthesizing a dataset of sounds using the functional transformation method.
Then, we represent each percussive sound in terms of its time-invariant scattering transform coefficients and formulate the parametric estimation of the resonator as multidimensional regression with a deep convolutional neural network.
We interpolate scattering coefficients over the surface of the drum as a surrogate for potentially missing data, and study the response of the neural network to interpolated samples.
Lastly, we resynthesize drum sounds from scattering coefficients, therefore paving the way towards a deep generative model of drum sounds whose latent variables are physically interpretable.
\end{abstract}
\section{Introduction}\label{sec:introduction}
Throughout musical traditions, drums come in all shapes and sizes (see Figure 1).
Such diversity in manufacturing results in a wide range of perceptual attributes: bright, warm, mellow, and so forth.
Yet, current approaches to drum music transcription, which are based on one-versus-all classification, fail to capture the multiple factors of variability underlying the timbre perception of percussive sounds \cite{8350302}.
Instead, they regard each item in the drum kit as a separate category, and rarely account for the effect of playing technique.
Therefore, in the context of music information retrieval (MIR), the goal of broadening and refining the vocabulary of percussive sound recognition systems requires to move away from discrete taxonomies.

\begin{figure}
 \centering
  \includegraphics[width=\linewidth]{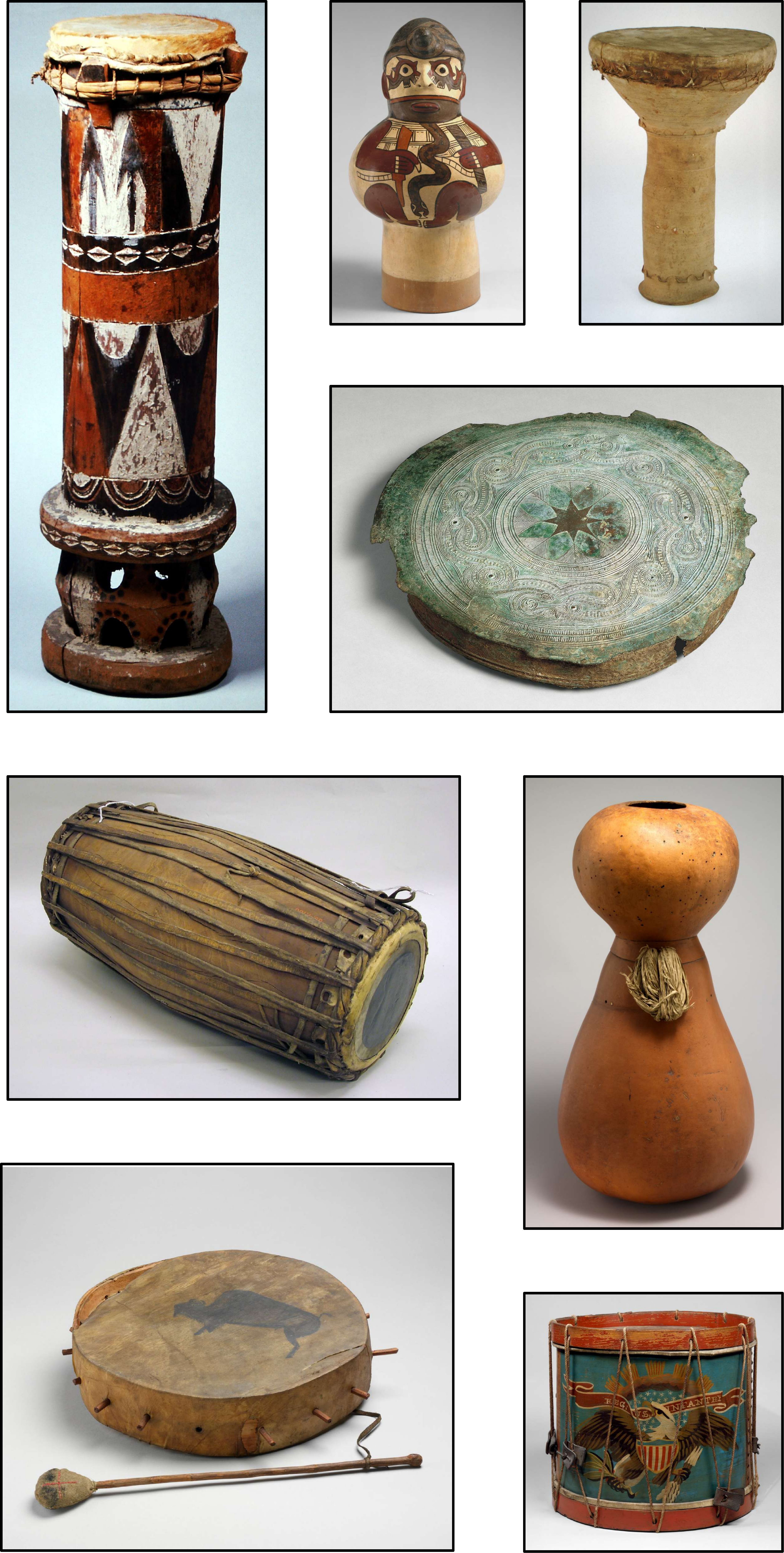}
  \caption{Drums of various shapes and materials.
  Left to right:
  mbejn, 19th century, Fang people of Gabon;
  ceramic drum, 1st century, Nasca people of Peru;
  darabukka, 19th century, Syria;
  tympanum of a Pejeng-type drum, Bronze age, Indonesia (Sumba);
  pakhavaj, 19th century, North India;
  ipu hula, 19th century, Hawai'i;
  frame drum, 19th century, Native American people of Dakota;
  Union army drum, ca. 1864, Pennsylvania.
  All images are in the public domain and can be accessed at: \protect\url{www.metmuseum.org}
  }
  \label{fig:drums}
\end{figure}

In a different context, prior literature on musical acoustics has managed to simulate the response of a drum from the knowledge of its shape and material.
Among studies on physical modeling of musical instruments, functional transformation method (FTM) \cite{trautmann_rabenstein_2013} and finite difference method (FDM) \cite{hiller1971synthesizing,antoine1994} play a central role.
They rely on partial differential equations (PDE) to describe the structural and material constraints imposed by the resonator.
The coefficients governing these equations may be varied continuously.
Thus, PDE-based models for drum sound synthesis offer a fine level of expressive control while guaranteeing physical plausibility and interpretability.

From a musical standpoint, a major appeal behind physical models lies in auditory perception: all other things being equal, larger drums tend to sound lower, stiffer drums tend to sound brighter, and so forth.
Yet, a major drawback of PDE-based modeling for drum sound synthesis is that all shape and material parameters must be known ahead of time.
If, on the contrary, these parameters are unknown, adjusting the synthesizer to match a predefined audio sample incurs a process of multidimensional trial and error, which is tedious and unscalable.
This is unlike other methods for audio synthesis, such as digital waveguide \cite{PASPWEB2010} or modal synthesis \cite{Adrien1991ms}.

In this article, we strive towards resolving the tradeoff between control and flexibility in drum sound synthesis.
To this end, we formulate the identification of percussive sounds as an inverse problem, thus combining insights from physical modeling and statistical machine learning.
Our main contribution is wav2shape, i.e., a machine listening system which takes a drum stroke recording as input and retrieves the shape parameters which produced it.
The methodological novelty of wav2shape lies in its hybrid architecture, combining feature engineering and feature learning: indeed, it composes a 1-D scattering transform and a deep convolutional network to learn the task of shape regression in a supervised way.
The advantage of choosing scattering coefficient over conventional audio descriptor such as MFCC and CQT in characterizing nonstationary sounds has been discussed in previous works \cite{vincent2018ipt,lostanlen2020eusipco}. 

The subtitle of this paper is a deliberate reference to a famous mathematical paper named ``Can One Hear the Shape of a Drum?'' \cite{kac1966amm}; that is, whether any two isospectral planar domains are necessarily isometric.
Since its publication, this question has been answered affirmatively in the important particular cases of circular and rectangular domains; but negatively in the general case, with the construction of nonconvex counterexamples.
Despite the evident connection with our paper, we note that \cite{kac1966amm} and wav2shape strive towards slightly different goals.
First, while \cite{kac1966amm} makes no prior assumption on the symmetries of the membrane, wav2shape focuses on representing rectangular and circular membranes, which are by far the most common in music.
In return, while \cite{kac1966amm} is restricted to the recovery of the domain under forced oscillations, wav2shape also expresses the effects of stiffness and damping, both frequency-dependent and frequency-independent.
These effects are crucial for modeling the response of the drum membrane to a localized impulse, e.g. induced by the player's hand, a stick, or a mallet.

Our main finding is that, after training, wav2shape is able to generalize to previously unseen shapes.
Add an additional experiment, we interpolate the value of scattering coefficients over the 2-D surface of the drum and verify that the convnet in wav2shape generalizes to interpolated drum stroke locations.
Lastly, we invert the scattering transform operator, thus laying the foundations for turning wav2shape into a deep generative model without explicit knowledge of the partial differential equation (PDE) underlying the vibration of the membrane.


\section{Problem Statement}\label{sec:Problem definition}

\subsection{Multidimensional regression of PDE coefficients}
The vibration of a drum obeys a partial differential equation (PDE), involving both resonant and dissipative terms.
In the following, we assume the analytical form, boundary conditions, and initial conditions of this PDE to be known---as Section \ref{sec:Model} will discuss.
Conversely, we take its vector of constant coefficients $\boldsymbol{\theta}$ to be unknown.
We represent the state of the drum by the displacement field $\mathbf{X}_{\boldsymbol{\theta}}$ of its membrane as a function of space \(u\in[0,l]\) and time $t$.
We place the origin of the Cartesian coordinate system at the center of the drum ($u=u_0=(l/2,l/2)$) and the onset of the stroke ($t=0$).
The goal of wav2shape is to recover $\boldsymbol{\theta}$ from a single measurement of  $\mathbf{X}_{\boldsymbol{\theta}}$ near the origin.

\subsection{Need for geometrical invariants}
Let $\bm{x}_{\boldsymbol{\theta}}:t \mapsto \mathbf{X}_{\boldsymbol{\theta}}\big(t, u=u_0\big)$ be the time series describing the displacement of the drum at its center.
For any given $\boldsymbol{\theta}$, the signal $\bm{x}_{\boldsymbol{\theta}}$ lasts for about one second and spans about \SI{20}{\kilo\hertz} in bandwidth.
Therefore, once discretized uniformly and truncated to a finite duration, $\bm{x}_{\boldsymbol{\theta}}$ has a typical length of $10^{5}$ samples.
Furthermore, Euclidean distances in the waveform domain are not informative for recovering $\boldsymbol{\theta}$: for example, flipping the polarity of the signal (i.e., from to $\bm{x}_{\boldsymbol{\theta}}$ to $-\bm{x}_{\boldsymbol{\theta}}$) produces a large Euclidean distance, yet leaves $\boldsymbol{\theta}$ unchanged.
More generally, discrepancies in audio acquisition across samples, e.g. involving changes in gain and DC bias, imply that the evolution of each $\bm{x}_{\boldsymbol{\theta}}$ is only known up to a global affine transformation.
For this reason, a major challenge underlying the development of wav2shape is to represent high-dimensional audio signals in a feature space which satisfies certain geometrical invariants (such as $\bm{x}_{\boldsymbol{\theta}}\mapsto a \bm{x}_{\boldsymbol{\theta}} + b$) while preserving informative variability in $\boldsymbol{\theta}$.

\subsection{Need for phase demodulation}
In addition to affine changes in the displacement domain, $\bm{x}_{\theta}$  is also subject to random fluctuations in the spatiotemporal domain.
This is because, in practice, the origin ($u=l/2$, $t=0$) of the Cartesian coordinate system is prone to small measurement errors.
Given that $\mathbf{X}_{\theta}$ oscillates rapidly in time and space near the origin, such measurement errors incur large phase deviations.
These phase deviations affect Euclidean distances between waveforms.
On the contrary, long-range interactions between wave ridges are informative of modal resonance and damping, regardless of phase.
Hence, wav2shape must demodulate fast oscillations in $\boldsymbol{x}_{\theta}$ in order to stably characterize shape parameters $\boldsymbol{\theta}$.

\subsection{Need for numerical stability to deformations}
Let us denote by $\boldsymbol{\Phi}$ an instance of the wav2shape model.
The output of $\boldsymbol{\Phi}$ is a vector of constant coefficients to the PDE governing the vibration of the drum: $\boldsymbol{\tilde{\theta}} =  \boldsymbol{\Phi}(\boldsymbol{x_{\theta}})$.
We evaluate wav2shape in terms of Euclidean distance between vectors describing true vs. predicted drum shapes:
\begin{equation}
    L_{\Phi} (\boldsymbol{\theta}) =
    \big\Vert \boldsymbol{\tilde{\theta}} - \boldsymbol{\theta} \big\Vert_2 =
    \big\Vert \boldsymbol{\Phi}(\boldsymbol{x_{\theta}}) - \boldsymbol{\theta} \big\Vert_2.
\end{equation}

This Euclidean distance is computed in a vector space of relatively high dimension---in this article, we encode $\boldsymbol{\theta}$ in dimension five.
Thus, the supervised prediction of $\boldsymbol{\theta}$ is exposed to the curse of dimensionality.
In order to learn the wav2shape function $\boldsymbol{\Phi}$ from limited annotated data, it is necessary to map waveform samples $\boldsymbol{x}_{\theta}$ to a feature space in which coordinate-wise variations of $\boldsymbol{\theta}$ are disentangled and linearized. 

In the context of wav2shape, some factors of variability in $\boldsymbol{\theta}$ (e.g., pitch) are most intuitive in the frequency domain, while others (e.g., rate of damping) are most intuitive in the time domain.
Therefore, it is advantageous to train a machine learning system to regress $\boldsymbol{\theta}$ in the time--frequency domain, rather than the time domain.
Section \ref{sec:Machine-listening} will present how wav2shape combines a scattering transform and a deep convolutional neural network, as an unsupervised feature extraction stage and a supervised nonlinear regression stage respectively.

\section{From shape to wave:\\physical synthesis model}\label{sec:Model}
\subsection{Formulation as a fourth-order PDE}
Let us recall the wave equation in dimension two:
\begin{equation}
    \dfrac{\partial^2 \mathbf{X}}{\partial t^2}
    - c^2
    \nabla^2\mathbf{X} = \mathbf{Y}(t, u),
\end{equation}
where $c$ is the speed of sound over the drum membrane; the symbol $\nabla^2$ denotes the spatial Laplacian operator; and the scalar field $\mathbf{Y}$ represents the gesture of the musician.
Throughout this article, we assume the spatiotemporal field $\mathbf{Y}$ to be factorizable into a temporal component $\boldsymbol{y}_{\mathrm{t}}$ and a spatial component $\boldsymbol{y}_{\mathrm{u}}$.

Although the formulation above may be sufficient to identify stationary eigenmodes in $\mathbf{X}$, it does not faithfully characterize the response of a drum membrane to a percussive excitation $\mathbf{Y}$ \cite{RABENSTEIN20031673}.
To address this issue, we consider the stiffness $S$ of the drum membrane as a function of its Young's modulus and its moment of inertia.
Furthermore, air drag induces an energy dissipation in $\mathbf{X}$ through a first-order damping coefficient $d_1$.
Lastly, near the boundary of the drum, the mechanical coupling between the membrane and the body of the drum also causes energy dissipation through a third-order damping coefficient $d_3$.

Once the terms $S$ (stiffness), $d_1$ (first-order damping), and $d_3$ (third-order damping) have been taken into account, the PDE governing the displacement field $\mathbf{X}$ becomes:
\begin{align}
    &\left(\dfrac{\partial^2 \mathbf{X}}{\partial t^2}(t,u)
    - c^2 \nabla^2\mathbf{X}(t,u) \right)
    \nonumber \\
    &+ S^4 \big(\nabla^4
    \mathbf{X}(t,u)\big)
    +\dfrac{\partial}{\partial t}
    \Big(d_1 \mathbf{X}(t,u) + d_3 \nabla^2\mathbf{X}(t,u)\Big) \nonumber \\
    &= \mathbf{Y}(t, u)
    = \boldsymbol{y}_{\mathrm{t}} (t) \boldsymbol{y}_{\mathrm{u}}(u),
    \label{eq:pde}
\end{align}
where the spatiotemporal field $\nabla^4 \mathbf{X}$ denotes the ``double Laplacian'' of $\mathbf{X}$, i.e., the Laplacian of $\nabla^2 \mathbf{X}$.

\subsection{Boundary conditions \label{sec:boundary}}
For the sake of simplicity and conciseness, we only address the case of a rectangular membrane, e.g., that of a caj\'{o}n.
The important case of a circular membrane (see Figure \ref{fig:drums}) could be derived from Equation \ref{eq:pde} with the same tools as presented hereafter; yet, it would incur a conversion to polar coordinates, and the resort to Bessel functions.
We direct readers to \cite{trautmann2001icassp} for the important case of the  circular membrane.
Note, in this case, that the transfer function method (TFM) is an alternative denomination for the functional transformation method (FTM).

We consider the membrane to be a rectangle of width $l_1$, length $l_2$, and aspect ratio $\alpha = l_1 / l_2$.
Along the edges of this rectangle, we assume the displacement field to be null: for every $t$, $\mathbf{X}(t, u) = 0$ if $u_1=0$, $u_1=l_1$, $u_2=0$, or $u_2=l_2$.
This is tantamount to assuming that the shape of the drum remains fixed throughout the duration of the percussive stroke.

\subsection{Functional transformation method (FTM)}
The Laplace transform of $\mathbf{X}$ over the time dimension is
\begin{equation}
    \mathcal{L}\{\mathbf{X}\} (s, u) =
    \int_{0}^{+\infty}
    \mathbf{X}(t, u)\,
    \exp(-st)\;\mathrm{d}t,
\end{equation}
In the Laplace domain, Equation \ref{eq:pde} becomes
\begin{align}
     &S^4 \left(\nabla^4\mathbf{\mathcal{L}\{X\}}(s,u)\right)&
     \nonumber \\
     +&\left(s d_{3}- c^2\right)\nabla^2\mathbf{\mathcal{L}\{X\}}(s,u)&
     \nonumber \\
     +&\left(s^2+sd_1\right)\mathbf{\mathcal{L}\{X\}}(s,u)&
     =\mathcal{L}\{\boldsymbol{y}_\mathrm{t}\}(s)
     \boldsymbol{y}_{\mathrm{u}}(u).
     \label{eq:pde_lap}
\end{align}

The interest of the Laplace domain is that, in comparison with Equation 3, the equation above replaces temporal derivatives with simpler algebraic terms. 
Similarly, spatial derivatives may be eliminated by means of the Sturm-Liouville transformation (SLT), as detailed in \cite{adjustable2016,8070610,5299084}.
Once in the Laplace-SLT domain, the solution of the PDE can be recovered in the spatiotemporal domain by performing an inverse Sturm-Liouville and inverse Laplace transform consecutively.
In this context, drums with a rectangular membrane are conceptually simpler: indeed, the inverse Sturm-Liouville transformation boils down to a Fourier series decomposition \cite{adjustable2016}.
Thus, in the particular case described in Section \ref{sec:boundary}, we may skip the SLT altogether and, instead, decompose the Laplace domain solution as a Fourier series over the 2-D variable $u$.

At any fixed $s\in\mathbb{C}$, the spatial field $u \mapsto \mathcal{L}\{\mathbf{X}\}(s,u)$ is absolutely continuous.
We index each mode in $\mathcal{L}\{\mathbf{X}\}$ by the pair $m = (m_1, m_2) \in \mathbb{Z}^2$, and denote by $\mathcal{\widehat{L}}_m(\mathbf{X})(s) \in \mathbb{C}$ the associated Fourier coefficients:
\begin{align}
    &\mathcal{L}\{\mathbf{X}\} (s,u)
    \nonumber\\
    &= \sum_{m\in\mathbb{N}^2}
    \mathcal{\widehat{L}}_m\{\mathbf{X}\}(s)
    \sin\left(\dfrac{m_1 \pi u_1}{l_1}\right)
    \sin\left(\dfrac{m_2 \pi u_2}{l_2}\right)
    \label{eq:x_lap}
\end{align}
Similarly, we decompose $\boldsymbol{y}^{\mathrm{u}}$ into a series of 2-D Fourier coefficients $\widehat{y}_{m}^{\mathrm{u}}$.
Plugging the equation above into Equation \ref{eq:pde_lap} allows a modal identification of the form:

\begin{align}
    \mathcal{\widehat{L}}_m\{\mathbf{X}\}(s)
    &=
    \mathcal{L}\{\boldsymbol{h}_m\}(s)
    \times
    \mathcal{L}\{\boldsymbol{y}^{\mathrm{t}}\}(s)
    \times
    \widehat{y}_{m}^{\mathrm{u}}
    \nonumber \\
    &=
    \frac{
    \mathcal{L}\{\boldsymbol{y}^{\mathrm{t}}\}(s)
    \times
    \widehat{y}_{m}^{\mathrm{u}}
    }{(s - z_m)(s - \overline{z_m})},
\end{align}
where the complex numbers $z_m$ and their conjugates $\overline{z_m}$ denote the poles of resonance of the impulse response $\boldsymbol{h}_m$.

After defining the constant $\gamma_{m} = m_1^2 + m_2^2 / \alpha^2$, a straightforward computation leads to
\begin{equation}
    \mathfrak{R}(z_m) =
    \dfrac{d_3 \gamma_m - d_1}{2}
    \label{eq:z_real}
\end{equation}
for the real part, and
\begin{equation}
    \mathfrak{I}(z_m)^2 =
    \left(
    S^2 - \dfrac{d_3^2}{4}
    \right)
    \gamma_{m}^2
    +
    \left(
    c^2 + \dfrac{d_1 d_3}{2}
    \right)
    \gamma_{m}
    -
    \dfrac{d_1^2}{4}
    \label{eq:z_imag}
\end{equation}
for the squared imaginary part.
Each impulse response $\boldsymbol{h}_m$ is a real-valued sine wave with an exponential decay:
\begin{equation}
    \boldsymbol{h}_m (t) =
    \exp\big(\mathfrak{R}(z_m)t\big)
    \sin\big(\mathfrak{I}(z_m)t\big).
    \label{eq:h}
\end{equation}
Lastly, an inverse Laplace transform of every term in Equation \ref{eq:x_lap} yields the following closed-form expression for $\mathbf{X}$:
\begin{align}
    \mathbf{X} (t, u)
    =
    &\sum_{m\in\mathbb{N}^2}
    \big(
    \boldsymbol{y}^\mathrm{t}
    \ast
    \boldsymbol{h}_m
    )(t)
    \nonumber \\
    &\times
    \widehat{y}_{m}^{\mathrm{u}}
    \sin\left(\dfrac{m_1 \pi u_1}{l_1}\right)
    \sin\left(\dfrac{m_2 \pi u_2}{l_2}\right),
\end{align}
where the asterisk denotes the convolution operator.

\subsection{Reparametrization}
Although the tuplet $(S, c, d_1, d_3, \alpha)$ suffices to describe the physical system in Equation \ref{eq:pde}, this tuplet remains unwieldy from a computer music standpoint.
Indeed, software plugins for drum sound synthesis usually have knobs for ``pitch'' and ``duration''; yet, these two perceptual attributes do not appear clearly in Equation \ref{eq:pde}.
Therefore, we map the tuplet above to a 5-D space in which pitch and duration may be controlled intuitively.

Given a mode $\boldsymbol{h}_m$ (see Equation \ref{eq:h}), we denote its carrier frequency by the imaginary part $\omega_m = \mathfrak{I}(z_m)$ and its modulation frequency by the negative real part $\sigma_m = -\mathfrak{R}(z_m)$.
The fundamental frequency of $\boldsymbol{h}_m$ is perceived as proportional to $\omega_m$ while its duration is perceived as inversely proportional to $\sigma_m$.
By convention, we take the mode of largest spatial extent as the reference for the fundamental frequency and duration of $\mathbf{X}$.
Setting $m = (1, 0)$ in Equation \ref{eq:z_imag} yields the fundamental frequency:
\begin{equation}
    \omega = \omega_{(1,0)} =
    \sqrt{
    \frac{\beta^4}{\alpha^2} S^4
    +
    \dfrac{\beta}{\alpha} c^2
    -
    \dfrac{1}{4}
    \left(
    \dfrac{\beta}{\alpha}
    d_3
    - d_1
    \right)^2},
\end{equation}
where the dimensionless constant $\beta = \alpha + 1/\alpha$ is associated to the aspect ratio $\alpha$ of the rectangular drum membrane (see Section \ref{sec:boundary}).
Finally, we define the duration of $\mathbf{X}$ as the inverse of the modulation frequency of the mode $\boldsymbol{h}_{(1,0)}$. Equation \ref{eq:z_real} becomes:
\begin{equation}
    \tau = \dfrac{1}{\sigma_{(1,0)}}
    =
    \dfrac{2}{d_1 - \dfrac{\beta}{\alpha} d_3}.
\end{equation}

Furthermore, we define the frequency-dependent damping of $\mathbf{X}$ as
\begin{equation}
    p = \dfrac{d_3}{\beta d_3 - \alpha d_1}
\end{equation}
and its dispersion as
\begin{equation}
    D = \dfrac{1}{\alpha \omega}
    \sqrt{S^4 - \dfrac{d_3^2}{4}}.
\end{equation}

We describe the ``shape'' of the drum as the 5-D vector $\boldsymbol{\theta} = (\omega, \tau, p, D, \alpha)$.
Once defined the value of $\boldsymbol{\theta}$, we iterate over the multiindex $m=(m_1, m_2)\in\mathbb{N}^2$, set $\gamma_{m} = m_1^2 + m_2^2 / \alpha^2$, and define the associated modulation frequency
\begin{equation}
    \sigma_{m} = \frac{1+p(\gamma_{m}-1)}{\tau}
\end{equation}
and squared carrier frequency
\begin{align}
    \omega_m^2
    &=
    D^2 \omega^2 \gamma_{m}^2
    \nonumber \\
    &+
    \left(
    \dfrac{(1-p)^2}{\tau^2}
    +
    \omega^2 (1-D^2)
    \right) \gamma_m
    \nonumber \\
    &- \dfrac{(1-p)^2}{\tau^2}.
\end{align}
Then we define the exponentially modulated sinusoid $\boldsymbol{h}_m : t \mapsto \exp(-\sigma_m t) \sin(\omega_m t)$ as in Equation \ref{eq:h}.
The infinite series $(\boldsymbol{h}_m)$ fully describes the response of the drum to an arbitrary excitation $\mathbf{Y}$ (see Equation \ref{eq:pde}).
In practice, we compute impulse responses $(\boldsymbol{h}_m)$ over a finite grid of $M^2=100$ modes, i.e., ten modes in each dimension.

Observe that the parameter $\tau$ affects only modulation frequencies $\sigma_m$ without affecting carrier frequencies $\omega_m$.
Conversely, the parameters $\omega$ and $D$ only affect carrier frequencies $\omega_m$ without affecting modulation frequencies $\sigma_m$.
As regards $p$ and $\alpha$, they affect both the carrier frequency and the modulation frequency of every mode.

\subsection{Real-time implementation as a VST plugin}
We implement the physical model above in the C++ language by means of the JUCE application framework\footnote{Link to download source code and executable binaries:\\ \url{https://github.com/lylyhan/Thesis}}.
As a result, our drum sound synthesizer is portable on Windows, Mac OS X, and Linux.
Furthermore, we package our software in the VST (Virtual Studio Technology) format.
Thus, it can be integrated into a digital audio workstation (DAW) such as Ableton Live, Adobe Audition, Audacity, Cubase, Logic Pro, Max/MSP, or REAPER.

\begin{figure}
\centering
\includegraphics[trim={0 0 0 15mm},clip, width=0.9\columnwidth]{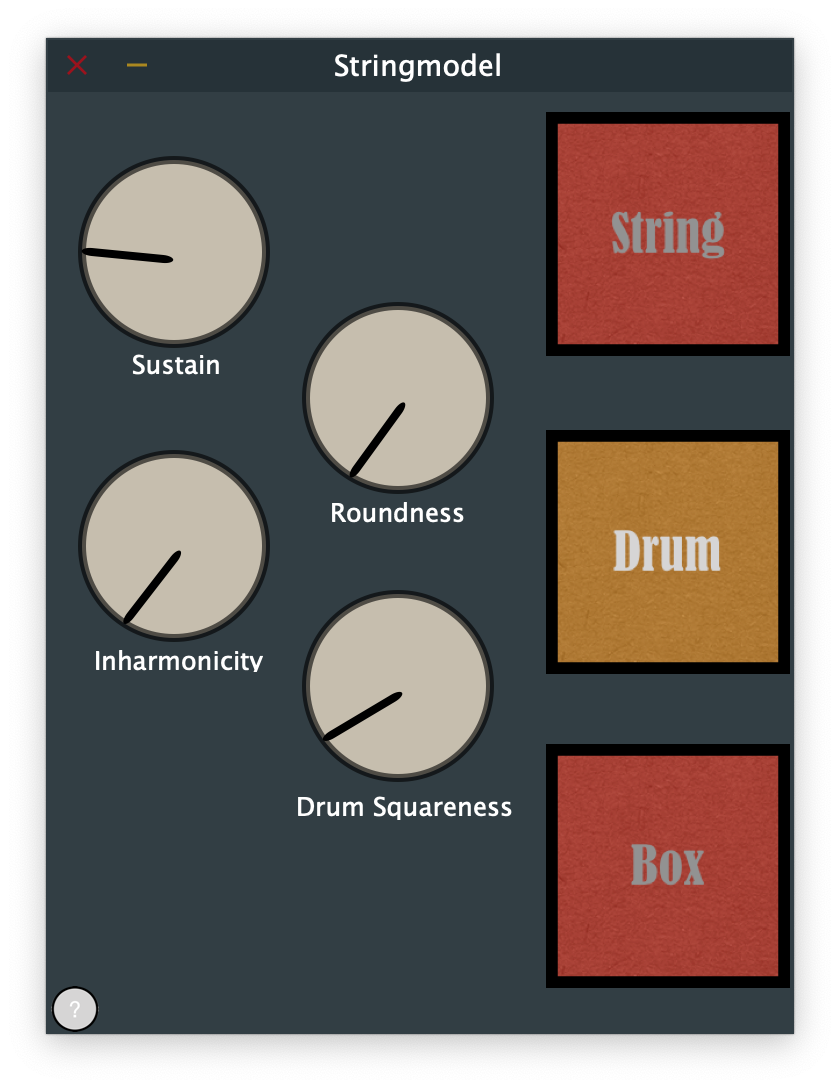}  
\caption{Screenshot of the graphical user interface (GUI) of our real-time drum synthesis plugin.}
\end{figure}

The plugin contains three physical models: 1D string model, 2D rectangular drum model, and 3D cuboid model.
All models solve the PDE in Equation \ref{eq:pde} via the functional transformation method (FTM), adapted to different dimensions of the spatial variable $u$.

Figure 2 illustrates the appearance of the plugin, in its ``drum'' (i.e., 2-D) variant.
The plugin contains four knobs, mapped as follows: ``sustain'' $\tau$, ``roundness'' $p$, inharmonicity $D$, and ``squareness'' (i.e., aspect ratio) $\alpha$.
Moreover, we map the logarithm of fundamental frequency $\omega$ to MIDI note numbers.
All knobs can be controlled by dragging the pointer on the screen or via a MIDI controller.

In practice, we set the excitation field $\mathbf{Y}$ to be a 2-D Gaussian of fixed width.
The user can set the center $u_0$ of this Gaussian by clicking inside a depiction of the drum membrane on the screen.
Our real-time implementation discretizes the time variable $t$ according to the chosen sampling rate, and renders the spatiotemporal field $\mathbf{X}_{\boldsymbol{\theta}}(t,u)$ by summing up modes in Equation in \ref{eq:h}.
Note that, as long as the shape vector $\boldsymbol{\theta}$ does not vary between drum hits, the values of $z_m$ and $\boldsymbol{h}_m$ can be precomputed without knowledge of $u_0$.

\section{From wave to shape:\\Machine listening model}\label{sec:Machine-listening}
Our problem statement (Section \ref{sec:Problem definition}) stressed the importance of geometrical invariants, phase demodulation, and numerical stability to deformations in the context of regressing shape ($\boldsymbol{\theta}$) from wave ($\boldsymbol{x}_{\boldsymbol{\theta}}$).
In this section, we present the ``wav2shape'' machine listening model and explain how it satisfies these mathematical properties.
This model has a hybrid architecture: it composes a feature engineering stage (1-D scattering transform) and a feature learning stage (deep convolutional network) in a supervised way.

\subsection{Scattering transform}




Let $\boldsymbol{\psi} \in \mathbf{L}^2(\mathbb{R}, \mathbb{C})$ a Hilbert-analytic filter with null average, unit center frequency, and quality factor $Q$ equal to one.
We define a wavelet filterbank as the family $\boldsymbol{\psi}_{j} : t \mapsto 2^{-j} \boldsymbol{\psi}(2^{-j} t)$ for integer $j$.
Each wavelet $\boldsymbol{\psi}_{j}$ has a center frequency proportional to $2^{-j}$ and an effective receptive field proportional to $2^{j}$ in the time domain.

We define the scalogram of $\bm{y}$ as the complex modulus of its discrete wavelet transform (DWT):
\begin{equation}
    \mathbf{U_1}\boldsymbol{x} : (t, j_1) \longmapsto
\left\vert \int_{-\infty}^{+\infty} \bm{x}(t^{\prime}) \bm{\psi_{j_1}}(t - t^{\prime})\;\mathrm{d}t^{\prime} \right \vert.
\label{eq:scalogram}
\end{equation}
Likewise, we define a second layer of nonlinear  transformation for $\bm{y}$ as the ``scalogram of its scalogram'':
\begin{equation}
\mathbf{U_2} \boldsymbol{x} : (t, j_1, j_2) \longmapsto
\Big\vert \big \vert \boldsymbol{x} \ast \bm{\psi_{j_1}} \big \vert \ast \bm{\psi_{j_2}} \Big \vert (t),
\end{equation}
where the asterisk denotes a convolution product.

Every layer in a scattering network composes an invariant linear system (namely, the complex DWT) and a pointwise operation (the complex modulus).
Thus, by recurrence over the depth variable $n$, every tensor $\mathbf{U_n} \bm{y}$ is equivariant to the action of delay operators.
This alternation of convolution and modulus transform provides complementary high-frequency wavelet coefficients \cite{Anden_2014}.

In order to replace this equivariance property by an invariance property, we integrate each  $\mathbf{U_n}$ over some predefined time scale $T = 2^J$, yielding the invariant scattering transform:
\begin{align}
    \mathbf{S_{n}} \bm{x} : (t, p) \longmapsto
    \int_{-\infty}^{+\infty} \mathbf{U_{n}}(t^{\prime}, p) \bm{\phi}_{T}(t-t^\prime)\;\mathrm{d}t^{\prime}
\end{align}
where the $n$-tuple $p = (j_1 \ldots j_n)$ is known as a scattering path and the function $\bm{\phi}_T$ is a real-valued low-pass filter of time scale $T$.
The number of layers is referred to as the order of the scattering transform.
Finally, we concatenate invariant scattering transform coefficients of different orders:
\begin{align}
    \mathbf{S} \bm{x}(t,p) = [\mathbf{S_{0}} \bm{x}(t), \mathbf{S_{1}} \bm{x}(t), ..., \mathbf{S_{N}} \bm{x}(t)](p),
\end{align}
where the path $p$ is a multiindex tuple containing between zero and $N$ entries.
We direct readers to \cite{mallat2012cpam} for further mathematical details on the scattering transform.

The two most important hyperparameters of the scattering transform are its scale $J$ and its order $N$.
A higher scale determines the window size, reduces the number of time bins, and produces more scattering coefficients for each time bin. 
Higher-order scattering coefficients encode and layer energy extracted from the maximum to a number of shorter time scales, and thus introduce a ``deep'', nonlinear characterization of spectrotemporal modulations.

In this article, we set $J=8$ and $N=2$ unless specified otherwise.
We compute the scattering transform by means of the Kymatio library \cite{andreux2020jmlr}, using PyTorch as a backend\footnote{Official website of Kymatio library: \url{https://kymat.io}}.



\subsection{Deep convolutional network: wav2shape}
In order to learn a nonlinear mapping between waveform and the set of physical parameters, we train a convolutional neural network, dubbed  wav2shape (``wave to shape'').
Comprising four 1-D convolutional layers and two fully connected dense layers, wav2shape is configured as follows:
\begin{itemize}
  \item layer 1: The input feature matrix passes through a batch normalization layer, then 16 convolutional filters with a receptive field of 8 temporal samples.
  The convolution is followed by a rectified linear unit (ReLU) and average pooling over 4 temporal samples.
  \item layer 2, 3, and 4: same as layer 1, except that the batch normalization happens after the convolution. The average pooling filter in layer 4 has a receptive field of 2 temporal samples, due to constraint in the time dimension. After that, layer 4 is followed by a ``flattening'' operation.
  \item layer 5: 64 hidden units, followed by a ReLU activation function.
  \item layer 6: 5 hidden units, followed by a linear activation function. 
\end{itemize}

Instead of supplying ``raw'' scattering coefficients to the first layer of wav2shape, we apply a logarithmic transformation of the form
\begin{equation}
\rho\big({\mathbf{S}}\boldsymbol{x}\big)(t,p) =
\log \left( 1+ \dfrac{\mathbf{S}\boldsymbol{x}(t,p)}{\varepsilon}\right),
\end{equation}
which has the effect of empirically Gaussianizing the statistical distribution of each coefficient \cite{vincent2018ipt}.

We set \(\varepsilon = 10^{-3}\) after verifying informally that this value yields features which are similar enough for slightly perturbed audio signals with imperceptible difference, yet still sufficiently distinct across different drum shapes $\boldsymbol{\theta}$.
Smaller values of the hyperparameter $\varepsilon$ yields more discriminating feature representations, however too small an $\varepsilon$ might magnify the otherwise imperceptible difference between audio signals in feature space.

During training, we minimize mean squared error between the ground truth and predicted \(\boldsymbol{\theta}\) using the Adam optimizer.
We use a minibatch size of 64 and train for 30 epochs with 50 steps per epoch, i.e. 96k samples in total.
The validation set accuracy is checkpointed after each epoch to identify the best performing model.

\section{Experiments}\label{sec:Experiments}

\subsection{Dataset}

We synthesize a dataset of percussive sounds by discretizing the physical parameters
\begin{equation}
    \boldsymbol{\theta} = \{\omega, \tau, \log p, \log D, \alpha\}
\end{equation}
uniformly, thus resulting in a five-dimensional hypercube.
Each sound is computed with the same temporal excitation; that is, a Dirac impulse in the time domain ($\boldsymbol{y}^\mathrm{t} = \delta_t$) and a Gaussian in the spatial domain: \begin{equation}
    \hat{y}_m^u=\mathit{G}(\mu=l/2,\sigma=0.4),
\end{equation}
peaking at the center of the drum.
Each drum sound lasts for $2^{15}$ audio samples, i.e., about $1.5$ second at a sample rate of $22050$ Hz.

Along each dimension of the five-dimensional hypercube of shape parameters \(\boldsymbol{\theta}\), we curate the validation set by carving out the ``center'' $60\%$ range over every dimension.
This results in \(0.6^5\approx 7.8\%\) of the total 100k samples being assigned to validation set.
Furthermore, we assign the surrounding sample space proportionally to training set and test set. 
This ensures that the training set, validation set, and test set do not overlap.
There are 82221 samples in training set, 10k samples in test set, and 7779 samples in validation set.

\subsection{Shape Regression}
\begin{figure}
\centering
\includegraphics[width=\columnwidth]{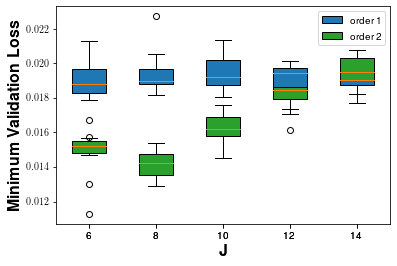}  
\includegraphics[width=\columnwidth]{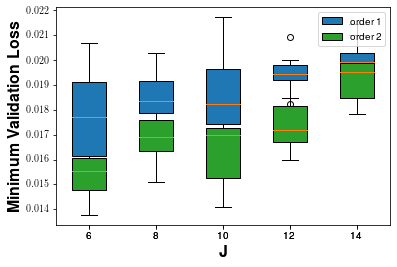}  
\label{fig:varyJ}
\includegraphics[width=\columnwidth]{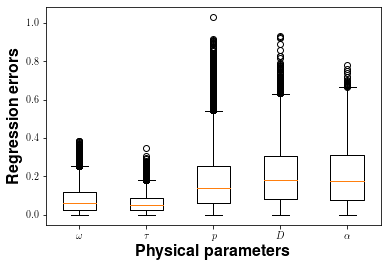}  
\label{fig:paramregress}

\caption{Training the convolutional neural network with different choices of scale $J$, order $N$ and scaling factor $\varepsilon$ as input scattering coefficients yields varying learning robustness. The upper 2 diagrams show effects on validation regression loss by selecting $J\in\{6,8,10,12,14\}$ and orders $N\in\{1,2\}$, where the upper and lower diagrams are using scaling factor of $\varepsilon \in \{10^{-3},10^{-1}\}$ respectively during feature preprocessing. With $\varepsilon=10^{-3}$, the best-performing model is $J=8$, order $N=2$. The bottom diagram demonstrates distribution of the absolute regression error of each individual physical parameter when applying the best model on test set. Box and whisker edges denote quartiles and deciles respectively.}
\end{figure}

The best performing wav2shape model results from a trial-and-error process of hyperparameter optimization.  
We perform ten trials of training with different values of scale $J$ and order $N$ to find the most successful input feature. 
Figure 3 (top) summarizes our findings.
Note that the input feature dimension varies with $J$: thus, when the resulting time dimension is small,the number of average-pooling filters and their receptive field sizes need to be changed accordingly.
Apart from the case of $J=8$ detailed in 4.2, $J=6$ uses four average-pooling of receptive field 4; $J=10$ uses two of receptive fields 4 and one of receptive field 2; $J=12$ uses one each of receptive field 4 and 2; and $J=14$ uses only one of receptive field 2.

We evaluate wav2shape in terms of Euclidean distance between prediction and the normalized ground truth \(\theta\). 
As a point of comparison, the mean Euclidean distance between two points drawn uniformly at random in a 5-dimensional hypercube of unit hypervolume is around 0.87.
In all of the models, the minimum validation loss is far below this value: this indicates that all variations of wav2shape generalize beyond the training set.
The best performing model is achieved with $J=8$ second-order scattering coefficients scaled by $\varepsilon=10^{-3}$ as input, where the lowest minimum validation loss across ten trials is around 0.0129.

Consistently with previous publications on the scattering transform, we observe that, for all values of $J$, shape regression with $N=2$ outperforms $N=1$.
Indeed, the double nonlinearity in second-order scattering transform contributes to the demodulation of nonstationarities, such as those found at the onset of a drum sound.
Meanwhile, larger scale $J$ increases the maximum time window size, thus encodes the audio signal $\boldsymbol{x}_{\boldsymbol{\theta}}$ with a lower sample rate yet more coefficients along the frequency dimension.
As $J$ increases, the audio descriptor is more stable to deformations yet less discriminative to variations in drum shape.

Figure 3 (bottom) breaks down the regression error of our best performing model according to different dimensions of the shape parameter \(\boldsymbol{\theta}\).
We observe that our model is the most accurate on parameters $\tau$ and  $\omega$ while being the least accurate on parameters $p$ and $D$.
An explanation is that \(\tau\) and \(\omega\) are the two parameters which more directly affect poles of the system.

On the other hand, both $D$ and $p$ have asymptotic influences on the poles as the mode number increases.
Specifically \(\tau_m \approx \tau_1/(pm^2)\) and \(\omega_m \approx D\omega_1m^2\) for large m.
These imply that effects of changing $p$ and $D$ would be more obvious when sound is synthesized with more modes. 
In our dataset each sound is summed only up to mode 10 due to time constraints. 
Thus this deficiency in higher modal data might have also caused the result.

\subsection{Hearing shapes from neighboring sounds}

\begin{figure}
    \centering
    \includegraphics[width=1.1\linewidth]{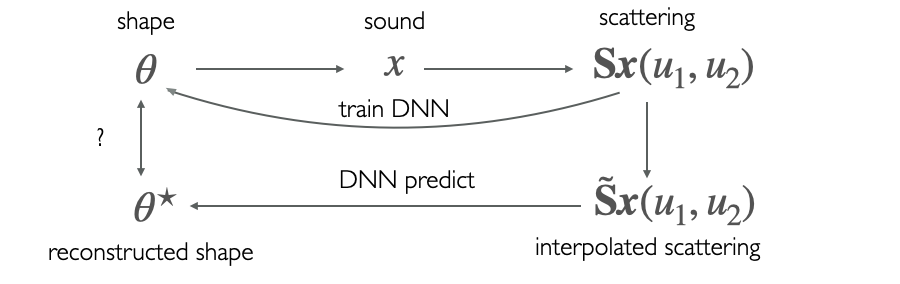}
    \caption{Diagram of the closed-loop system: we use physical sound synthesis, scattering transform, gradient descent, and the trained deep neural network wav2shape to traverse between physical parameters, sound and scattering domain. DNN: Deep Neural Network.  GD: Gradient Descent.}
    \label{fig:closed}
\end{figure}

To examine the stability of scattering transform, we construct a closed-loop system that allows us to traverse between sound, physical and scattering domains.
Figure \ref{fig:closed} shows a diagram of this closed-loop system.

We begin by selecting a drum shape, i.e. some random combination of physical parameters \(\boldsymbol{\theta}\). 
Secondly, we interpolate scattering transform coefficients \( \mathbf{SX} (t, u_1,u_2)\) on the drum.
Specifically, we compute the scattering transform at its neighboring points: $\mathbf{SX} (t,u_1-\delta,u_2)$, $\mathbf{SX} (t,u_1+\delta,u_2)$, $\mathbf{SX} (t,u_1,u_2-\delta)$, $\mathbf{SX} (t,u_1,u_2+\delta)$ and approximate the scattering coefficients at $(u_1,u_2)$ from those of its neighbors.
Thirdly, we regress physical parameters from  \(\widetilde{\mathbf{S}}\bm{x}(u_1,u_2)\) via the wav2shape model, yielding a vector \(\boldsymbol{\theta}^\star\).
Lastly, we measure the mean squared error between the predicted shape \(\boldsymbol{\theta}^\star\) and the true shape \(\boldsymbol{\theta}\).

The motivation behind this interpolation procedure is two-fold.
First, we examine the ability of the scattering transform to linearize the dependency of the drum signal $\boldsymbol{x}_{\theta}$ with respect to the location of the stroke.
Secondly, we inquire whether wav2shape, which is trained on signals measured at the exact center of the drum, remains capable of predicting the shape from surrounding measurements.
 
In order to approximate the scattering coefficients at $(u_1,u_2)$, we apply a four-point linear interpolation, i.e., an unweighted average of neighboring coefficients along the four cardinal directions: North, East, South, and West.
We measure the approximation error of each scattering path $p$ in terms of its discretized Laplacian
\begin{align}
\nabla^2 \mathbf{S}\mathbf{X}_{\boldsymbol{\theta}}(t, u_1,u_2, p)
&= \mathbf{S}\mathbf{X}_{\boldsymbol{\theta}}(t, u_1,u_2, p) \nonumber \\
&-\frac{1}{4} \mathbf{S}\mathbf{X}_{\boldsymbol{\theta}}(t,u_1-\delta,u_2,p) \nonumber \\
&-\frac{1}{4} \mathbf{S}\mathbf{X}_{\boldsymbol{\theta}}(t,u_1+\delta,u_2,p) \nonumber \\
&-\frac{1}{4} \mathbf{S}\mathbf{X}_{\boldsymbol{\theta}}(t,u_1,u_2-\delta,p) \nonumber \\
&-\frac{1}{4} \mathbf{S}\mathbf{X}_{\boldsymbol{\theta}}(t,u_1,u_2+\delta,p),
\end{align}
where the step size $\delta$ is equal to $10\%$ of the side length of the drum.
For a given scattering path $p$ and time instant $t$, the equation above measures the curvature of the manifold associated to $u\mapsto\mathbf{S}\mathbf{X}_{\boldsymbol{\theta}}(t,u,p)$.
If this manifold is approximately flat, the linear interpolation is relatively accurate and the discretized Laplacian is relatively small.

We summarize the discretized Laplacian above by taking its $\ell^2$ norm over time and across scattering paths:
\begin{align}
    \mathbf{H}\mathbf{X_{\boldsymbol{\theta}}}(
    &u_1, u_2)=
    \nonumber \\
    &
    \sqrt{
    \int_{\mathbb{R}}
    \sum_{p}
    \nabla^2 \mathbf{S}\mathbf{X}_{\boldsymbol{\theta}}(t, u_1,u_2, p)^2
    \;\mathrm{d}t,
    }
\end{align}
thus yielding a matrix which is indexed by the spatial variable $u = (u_1, u_2)$.

As an illustration, Figure 5 (left) shows the matrix $\mathbf{H}\mathbf{X_{\boldsymbol{\theta}}}$ as a heatmap, for a fixed value of the vector $\boldsymbol{\theta}$.
As a point of comparison, we also compute a Laplacian heatmap for Fourier modulus coefficients (Figure 5, right).
We observe a symmetric pattern over the surface of the drum.
The darker regions of this pattern correspond to the locations on the drum in which the approximation of scattering coefficients by means of linear approximation is the least valid.
Interestingly, the locations of best fit do not lie near the center, but between the four axes of symmetry of the drum.

By application of the Parseval theorem, the scattering transform and the Fourier transform have comparable $\ell^2$ norms, i.e., the norm of the signal $\boldsymbol{x}_{\boldsymbol{\theta}}$ \cite{mallat2012cpam}.
Therefore, the heatmaps in Figure 5 can be compared with the same numerical graduations.
Over the  surface of the drum, we observe that the Laplacian of the scattering transform has a smaller $\ell^2$ norm than the Laplacian of the Fourier transform modulus.
This difference reflects the better ability of the scattering transform to linearize the dependency of the signal $\boldsymbol{x}_{\boldsymbol{\theta}}$ with respect to the origin $u$ of the excitation.

\begin{figure}
\centering
\includegraphics[width=0.45\columnwidth]{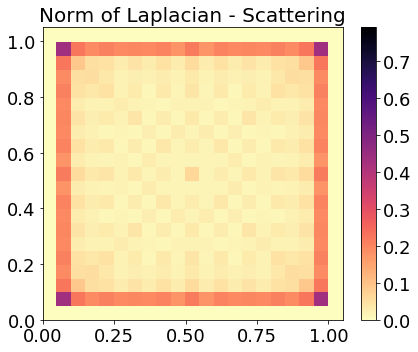}  
\includegraphics[width=0.45\columnwidth]{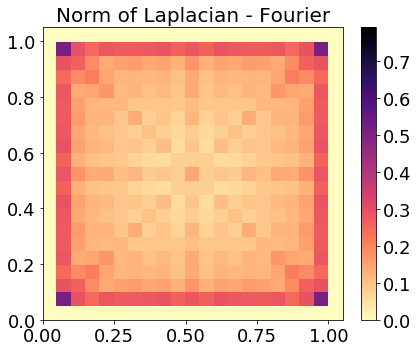}  
\includegraphics[width=0.45\columnwidth]{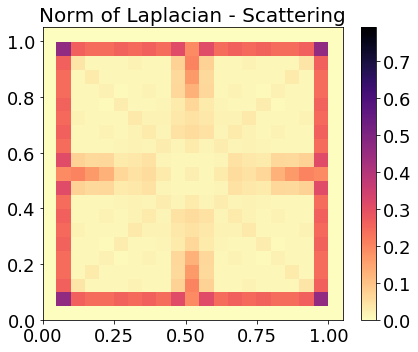}  
\includegraphics[width=0.45\columnwidth]{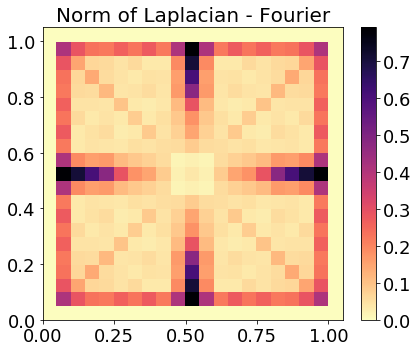} 
\label{fig:heatmap}
\caption{Heatmaps of Laplacian of scattering coefficients (left) and Fourier coefficents (right) on the drum. The upper pair uses Dirac impulses to model the excitation in space and the lower pair uses a Gaussian-shaped spatial envelope. Darker colors reflect a larger $\ell^2$ norm of the Laplacian.}
\end{figure}


As an additional experiment, we apply the wav2shape model to interpolated scattering coefficients.
We sample the shape vector $\theta$ from three distinct distributions: the validation set (7779 samples), the test set (10k samples with same distribution as training set), and a previously unseen test set (10k samples) drawn uniformly at random.

Figure 6 summarizes our results.
We find that wav2shape is capable of recovering the shape vector $\boldsymbol{\theta}$ with a relative mean squared error around $0.15$.
In comparison, a random guess would yield a relative mean squared error of the order of $0.87$ (see Section 5.2).
However, the error of wav2shape on interpolated scattering coefficients is larger than the error on true scattering coefficients, i.e., 0.0129 on the validation set.
Such discrepancy in shape regression accuracy results from interpolation error, manifested by the nonzero Laplacian at $u=l/2$ on the drum (see Figure 5). 
Future work will investigate methods to improve the ability of wav2shape to generalize to off-center stroke locations.

\begin{figure}
\centering
\includegraphics[width=\columnwidth]{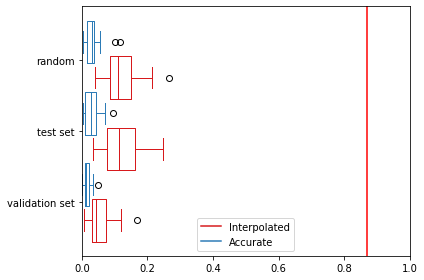}  
\caption{Comparison of prediction error when the best performing wav2shape model is applied onto scattering coefficients that are synthesized versus interpolated at $u=0$ on the same drum. 60 drums are randomly selected from three distributions: validation set (unseen by the model), test set (same distribution as training set) and random range (unseen by the model). The red line at 0.87 indicates regression loss achieved by a uniform random guess.}
\end{figure}

\subsection{Reconstruction}

\begin{figure*}
\centering
\includegraphics[width=1.9\columnwidth]{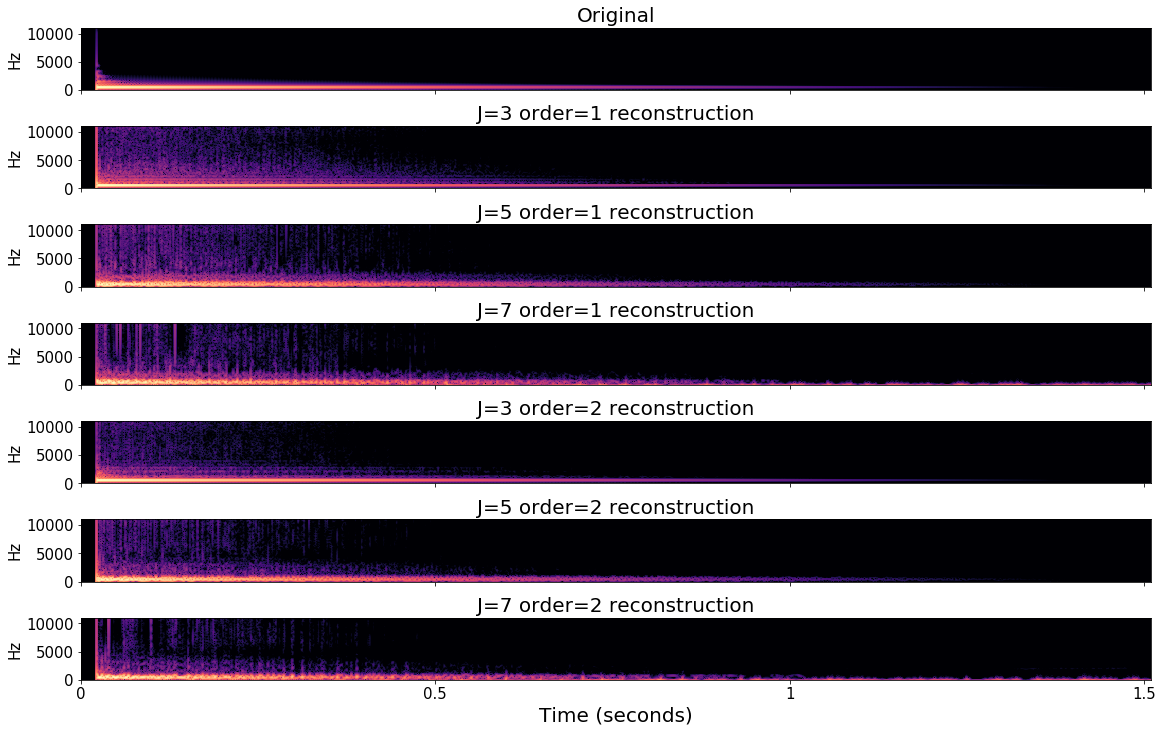}
\caption{Top plot: spectrogram of a synthetic drum sound. Bottom six plots: spectrograms of reconstructed sounds from order 1 and order 2 scattering coefficients, with different valus of the maximum scattering scale: $J\in\{3, 5, 7\}$.
Brighter colors denote greater values of energy in the time--frequency domain.}
\end{figure*}

In this section, we propose to invert the scattering transform operator; that is, for a given shape vector $\boldsymbol{\theta}$, to recover the drum sound $\boldsymbol{x}_{\boldsymbol{\theta}}$ from its scattering coefficients $\mathbf{S}\boldsymbol{x}_{\boldsymbol{\theta}}$.
The long-term motivation behind this approach is to convert wav2shape, which is a discriminative model, into a generative model.
Indeed, although the synthesis of drum sounds from simple shapes can be resolved by solving a partial differential equation in closed form (see Section 3), the case of irregular shapes is mathematically intractable.
One way to circumvent this problem is to adopt a data-driven perspective on physical simulations, and let a machine learning model decode the latent vector $\boldsymbol{\theta}$ into its acoustical correlate $\boldsymbol{x}_{\boldsymbol{\theta}}$.

Audio texture synthesis from scattering coefficients was inaugurated by \cite{bruna2013arxiv}.
We refer to \cite{lostanlen2019chapter} for a practical application of scattering-based audio synthesis to contemporary music creation.
More recently, the combination of scattering transform and generative adversarial networks (GAN) has shown a promising avenue of research \cite{andreux2018ismir,angles2018iclr}.

Given a drum shape vector $\boldsymbol{\theta}$, our goal is to minimize the error functional
\begin{equation}
    E(\boldsymbol{y}) =
    \big \Vert
    \mathbf{S}\boldsymbol{y}
    -
    \mathbf{S}\boldsymbol{x}_{\boldsymbol{\theta}}
    \big \Vert,
\end{equation}
defined as a Euclidean distance in the domain of scattering coefficients.
Let us denote by $\boldsymbol{\nabla}E(\boldsymbol{y})$ the gradient of $E$ evaluated at $\boldsymbol{y}$.
In practice, we compute $\boldsymbol{\nabla}E(\boldsymbol{y})$ by reverse-mode automatic differentiation, via the PyTorch backend of the Kymatio library.
We refer to \cite{vincent2019dafx} for a description of gradient backpropagation in a scattering network, with an application to audio texture synthesis.

Because the function $E$ is nonconvex, gradient descent does not converge to the global minimizer $\boldsymbol{y}^* = \boldsymbol{x}_{\boldsymbol{\theta}}$, but merely to a local minimizer.
Whether this local minimizer is perceptually to the target signal $\boldsymbol{x}_{\boldsymbol{\theta}}$ depends upon the parameters of the scattering transform.

Starting from a colored Gaussian noise $\boldsymbol{y}_0 (t)$ whose power spectral density matches $\mathbf{S}_1 \boldsymbol{x} (t,\lambda)$, we refine it by additive updates of the form \begin{equation}
    \boldsymbol{y}_{n+1}(t) = \boldsymbol{y}_n (t) + \boldsymbol{u}_n (t).
\end{equation}
In the simplest form of gradient descent, the signal $\boldsymbol{u}_n$ is equal to the gradient  $\boldsymbol{\nabla}E(\boldsymbol{y_n})$ multiplied by a constant learning rate term $\mu$.

To accelerate convergence, we adopt an adaptive learning rate policy, known as the ``bold driver'' heuristic \cite{sutskever2013icml}.
We initialize the learning rate at $\mu_0 = 0.1$.
At every iteration $n$, if the error has reduced, we confirm the update and increase the learning rate by $10\%$; otherwise, we retract the step and decrease the learning rate by $50\%$.
This leads to a sequence of learning rates: $\mu_1, \mu_2, $ etc. which depends upon the iteration.

Moreover, we add a momentum term to the gradient term $\boldsymbol{\nabla}E(\boldsymbol{y_n})$, leading to an update of the form:
\begin{equation}
    \boldsymbol{u}_n (t) = m \times \boldsymbol{u}_{n-1} (t) + \mu_n \boldsymbol{\nabla}E(\boldsymbol{y}_n)(t).
\end{equation}
Following a previous publication \cite[Section 3.3.2]{lostanlen2017phd}, we set the momentum hyperparameter to $m=0.9$.

After $n\sim 300$ iterations, the reconstruction error $E(\boldsymbol{y}_n)$ is two orders of magnitude below the initial reconstruction error $E(\boldsymbol{y}_0)$.
We repeat the same procedure for scattering networks of depths $N=1$ and $N=2$ and for various values of the scattering scale parameter $J$.

Figure 7 illustrates our results for one particular setting of the drum shape vector $\boldsymbol{\theta}$.
In the case of first-order scattering, we find that lower values of $J$ lead to a sharper reconstruction of $\boldsymbol{x}_{\boldsymbol{\theta}}$, at the expense of stability to deformations (see Section 2.4).
Conversely, larger values of $J$ elicit audible artifacts in the reconstructed signal $\boldsymbol{y}$.
Nevertheless, increasing the depth of the scattering network from $N=1$ to $N=2$ layers yields an improvement of reconstruction quality, which is particularly noticeable with higher values of $J$.
Although previous publications had reported the same effect in the case of audio textures \cite{bruna2013arxiv}, such as applause or bubbling water, the originality of our finding is that it applies to isolated percussive events.

\section{Conclusions}
In this article, we have presented wav2shape: a convolutional neural network which disentangles and retrieves physical parameters from waveforms of percussive sounds.
First, we have presented a 2-D physical model of a rectangular membrane, based on a fourth-order partial differential equation (PDE) in time and space.
We have solved the PDE in closed form by means of the functional transformation method (FTM), and included a freely downloadable VST plugin which synthesizes drum sounds in real time.
Then, we have computed second-order scattering coefficients of these sounds and designed wav2shape as a convolutional neural network (CNN) operating on the logarithm of these coefficients.
We have trained wav2shape in a supervised fashion in order to regress the parameters underlying the PDE, such as pitch, sustain, and inharmonicity.

From an experimental standpoint, we have found that wav2shape is capable of generalizing beyond its training set and predicting the shape of previously unseen sounds (Figure 3).
The network's robustness in shape regression confirmed that the scattering transform has the ability to linearize the dependency of the signal upon the position of the drum stroke (Figure 5).
Indeed, when applied on linearly interpolated scattering coefficients, the wav2shape neural network continues to produce an interpretable outcome.
Lastly, we have used reverse-mode automatic differentiation in the Kymatio library to synthesize drum sounds directly from scattering coefficients, without explicitly solving a partial differential equation.

Although the results of wav2shape are promising, we acknowledge that it suffers from some practical limitations, which hamper its usability in computer music creation.
First, physical parameters such as inharmonicity $D$ and aspect ratio $\alpha$ are not recovered as accurately as pitch $\omega$ or sustain $\tau$.
Secondly, wav2shape is only capable of retrieving the shape vector $\boldsymbol{\theta}$ if the rectangular drum is stroked exactly at its center: it would be beneficial, albeit challenging, to generalize the approach to any stroke location $u_0$.
Thirdly, we have trained wav2shape on a relatively large training set of over 82k audio samples.
The acquisition of these samples was only made possible by simulating the response of the membrane.
The prospect of extending autonomous systems from such a simulated environment towards a real environment is a topic of ongoing research in reinforcement learning, known as sim2real \cite{chebotar2019icra}.
Yet, the field of deep learning for musical acoustics predominantly relies on supervised learning techniques instead of reinforcement learning.
In this context, we believe that future research is needed to strengthen the interoperability between physical modeling and data-driven modeling of musical sounds.

\section{Acknowledgment}
This work is partially supported by National Science Foundation award 1633259 (BIRDVOX).
We wish to thank Jennie Choi for managing the open-access API of the Metropolitan Museum of Art and Ivan Selesnick for generously providing his notes on digital sound synthesis using the functional transformation method.
We also thank Scott Fitzgerald, Amy Hurst, and Mark Plumbley for fruitful discussions.

\clearpage
\bibliography{fa2020_template}

%
%
%

\end{document}